\documentclass[a4paper,12pt]{article}
\usepackage[T1]{fontenc}
\pdfoutput=1 

\usepackage{jheppub}
\usepackage{macro}
\usepackage{xcolor, tikz, pgfplots}
\usetikzlibrary{matrix,calc,positioning,decorations.markings,decorations.pathmorphing,decorations.pathreplacing}
\usetikzlibrary{arrows,cd,shapes}

\title{\textbf Soliton solutions and gravitational solitons: an overview}

\author[a]{Federico Manzoni}

\affiliation[a]{Mathematics and Physics department, Roma Tre}

\emailAdd{federico.manzoni@uniroma3.it}

\abstract{In this work we intend to discuss the solitonic solutions of Einstein's field equations in vacuum by constructing the solution to $N$ solitons and studying some aspects of it. In conclusion, it will be shown how the Kerr black hole can be interpreted as a soliton solution in the case of axial symmetry.}

\begin{document}

\maketitle
\section{Introduction}
Solitons were first observed in 1834 by the Scottish engineer John Scott Russell in the Union Canal; later, with the passage of time, the solitons were observed in other contexts such as, for example, in oceanography. By soliton we mean a particular wave-like solution of non-linear differential equations in which the dispersion effect is exactly balanced by the non-linearity effects; as we will see, an important characteristic of solitonic solutions is that in their interaction they re-emerge unchanged except for a phase shift.
With the intensification of research in the field of non-linear mathematical physics and on the theory of solitons, it has been discovered that this type of solution is present in many equations of great physical interest: from fluid mechanics to non-linear optics, from physics of superconductors to lattice defects in crystals, and in many other areas of physics and beyond\footnote{As, for example, in the study of telecommunications and DNA.}.
In fluid mechanics, the Korteweg-de Vries (KdV) equation describes long surface waves\footnote{By long wave we mean that $ hk << 1 $, where $ h $ is the depth of the seabed and $ k $ the wave number of the water wave; example of long waves are the offshore tsunamis. If instead $ hk >> 1 $ then we speak of short waves.} in the one-dimensional approximation and it admits soliton solutions.
In the context of non-linear optics the non-linear Schroedinger equation (NLS) derives from the Helmholtz equation once it is assumed that the variation of the refractive index of the material is small compared to the refractive index itself, that it depends only by the intensity of the light and that there is paraxiality. The NLS admits solutions of solitonic character.
In superconductor physics and crystallography, the sine-Gordon equation describes the dynamics, respectively, of the quantized magnetic flux in a Josephson junction\footnote{The Josephson junction is formed by two superconducting plates interspersed by an insulating membrane and is based on the effect tunnel of Cooper pairs (bonded states of two electrons or holes).} and of crystal dislocations; also the sine-Gordon equation has soliton like solutions.
\\
The main method for the search for solitonic solutions, the spectral transform, will be discussed in section \ref{ch2} together with the Darboux transform, another important mathematical tool of the soliton theory that allows us to build solitonic solutions knowing a particular solution of our problem. These tools will then be applied briefly to the case of the KdV showing, in this specific case, some general considerations made in paragraph \ref{sub2.1}. We will mainly refer to the bibliographic references \cite{7},\cite{4} and \cite{11}. \\
In this paper we will mainly focus on the solitons that emerge from Einstein's field equations in vacuum. In section \ref{ch3} we will discuss the generalization of the spectral transform to the theory of gravitation and, in addition to constructing the solution to $ N $ solitons and studying some properties of gravitational solitons, it will be shown how the Kerr black hole can be interpreted as a gravitational double soliton solution.
\subsubsection {Notations}
This conventions are adopted:
\begin{itemize}
    \item derivatives are indicated with commas, for example $f_{i, z, z}=\partial_z^2f_i =\frac{\partial^2f_i}{\partial z^2} $;
    \item identity matrix will be indicated as $\mathcal{I}$;
    \item traces are indicated with a capital letter, for example $R=Tr(\hat{R}) =Tr(R_ {\alpha \beta})$;
    \item The commutator is denoted as $ [\cdot, \cdot] $;
    \item The indices $i,j,h $ are to be considered added if repeated.
\end{itemize}

\section{Nonlinear equations and soliton solutions}\label{ch2}

As already anticipated in the introduction, many nonlinear equations of physical interest show soliton solutions. Therefore, let us concentrate on solving nonlinear equations; the methods we will develop are the spectral transform and the Darboux transform, which will be discussed in their general features and applied to the specific case of KdV. In the first paragraph we will mainly refer \cite{11},\cite{4} and \cite{7}.

\subsection{Spectral transform and Darboux transform}\label{sub2.1}
The spectral transform (TS) method was introduced in 1967 by Gardner, Greene, Miura and Kruskal \cite{2} to solve the problem at the initial values of the KdV and then extended to other situations of interest, for example as did Zakharov and Shabat in 1972 \cite{8} with the NLS. The method is based on the possibility of writing the equation under consideration as a condition of integrability of an associated matrix system of two linear differential equations, the so-called the Lax pair of the problem. This has the form
\begin{equation}
\begin{aligned}
\boldsymbol{\psi}_{,z}=\hat{U}\boldsymbol{\psi},\\
\boldsymbol{\psi}_{,t}=\hat{V}\boldsymbol{\psi},
\end{aligned}
\end{equation}
where $\boldsymbol{\psi}$, $\hat{U}$ and $\hat{V}$ depend on a parameter, $\lambda $, called the spectral parameter and on the variables $z$ and $t$. The spectral parameter satisfies the isospectrality condition\footnote{This can be shown in a simple way for the case of KdV, as we will see later.} $\lambda_{, t}=0 $. The condition of integrability is obtained as
\begin{equation*}
    \boldsymbol{\psi}_{,t,z}=\boldsymbol{\psi}_{,z,t} \Rightarrow \hat{V}_{,z}\boldsymbol{\psi}+\hat{V}\boldsymbol{\psi}_{,z}=\hat{U}_{,t}\boldsymbol{\psi}+\hat{U}\boldsymbol{\psi}_{,t} \Rightarrow \hat{V}_{,z}\boldsymbol{\psi}+\hat{V}\hat{U}\boldsymbol{\psi}=\hat{U}_{,t}\boldsymbol{\psi}+\hat{U}\hat{V}\boldsymbol{\psi},
\end{equation*}
from which follows the operatorial equation 
\begin{equation}
    \hat{V}_{,z}+\hat{V}\hat{U}-\hat{U}_{,t}-\hat{U}\hat{V}=0 \Rightarrow \hat{V}_{,z}-\hat{U}_{,t}+[\hat{V},\hat{U}]=0,
\end{equation}
said the Lax equation corresponding to the original equation of our Cauchy problem.
Imagine that our nonlinear differential equation determines the quantity $u(z,t)$ and that the initial condition $ u(z,0)$ is given. Once the appropriate Lax pair has been determined, the problem is divided into two parts: the direct problem and the inverse problem. The direct problem consists in associating the spectral parameters, such as the transmission coefficient $T(\lambda)$, the reflection coefficient $R(\lambda)$ and the elements $\lambda_n$ of the discrete spectrum, which correspond to the potential $u(z,0)$; these spectral parameters are denoted, as a whole, with $S(\lambda,0)$. The inverse problem, on the other hand, consists in reconstructing the potential at a generic time, $u(z,t)$, given the spectral parameters at time $t$, obtained by solving the spectral problem given by the Lax pair. The evolved spectral parameters are indicated with $S(\lambda,t)$.
In summary, the idea is therefore not to find the solution $u(z,t)$ directly from the starting equation because this evolution is often too complicated, therefore it is preferable to solve the associated spectral problem, which gives us the evolution over time of the spectral parameters and from the evolved spectral parameters we can obtain the solution $u(z,t)$. The fact that it is preferable derives from the linearity of the problem when looking at it from the point of view of the spectral problem; in fact, the evolution of the spectral parameters is ultimately given by the Lax pair, which is a matrix but linear differential system. The TS method is shown in the diagram below in Figure \ref{TS}. \\
From Figure \ref{TS} we note a certain degree of analogy with the Fourier transform scheme: in fact, the spectral transform method can be considered as a generalization to non-linear equations of the Fourier method for linear equations, and it reduces to this in the small field limit, $|u(z,t)|<<1$, and purely continuous spectrum.
\begin{equation}
\begin{aligned}
& \ \ \ \ \ \ \ \ \ \ \ \ \ \ \ \ \ \ \ \ \  \ \ \ \ u(z,0) \xrightarrow{direct \ problem} S(\lambda,0)\\
&non \ linear \ evolution \Bigg\downarrow\ \ \ \ \ \ \ \ \ \ \ \ \ \ \ \ \ \ \ \ \ \ \ \   \Bigg\downarrow \ linear \ evolution \\
&\ \ \ \ \ \ \ \ \ \ \ \ \ \ \ \ \ \ \ \ \  \ \ \ \ u(z,t) \xrightarrow{inverse \ problem } S(\lambda,t)
\label{TS}
\end{aligned}    
\end{equation}
We are interested in solitonic solutions, which emerge in the presence of a potential $u(z,0)$ corresponding to $R(\lambda)=0$; in this case the inverse problem is considerably simplified since we pass from having to solve an integral equation to an algebraic system. The TS method shows us how solitons are intimately related to the discrete elements $\lambda_n$ of the spectrum: for each of them a soliton emerges; moreover each $\lambda_n$ is a pole of the transmission coefficient and this implies that there are as many solitons as there are poles of $T(\lambda)$.

We now introduce another instrument of particular importance: the Darboux transform (TD). First, it is important to underline that TD is an algebraic method; it consists in constructing exact solutions for a non-linear integrable equation\footnote{It is intended to be integrable with the TS method. There is therefore a Lax pair whose compatibility condition is the starting equation. In the case of the KdV, and not only, this leads to an infinite number of conservation laws. See \cite{2} or \cite{4}}, Starting from the knowledge of a simple exact solution called background solution, that is
\begin{equation}
    \boldsymbol{\psi}=\hat{\chi}\boldsymbol{\psi}_b;
\end{equation}
the $\hat{\chi}$ matrix is called the dressing matrix. Assuming a dressing matrix of the shape
\begin{equation}
\hat{\chi}=\mathcal{I}+\sum_{k=1}^N\frac{\hat{O}_k}{\lambda-\lambda_k}
\end{equation}
in which the $\hat{O}$ operator does not depend on the spectral parameter, the $N$ solitons solution is constructed: a $N$ poles dressing matrix can generate the $N$ solitons solution. \\
The TD and the TS are linked by the fact that, as will be explicitly seen in the case of the KdV, the TD adds a pole to the transmission coefficient and this involves the addition of a soliton.

\subsection{The KdV case}
In the previous paragraph the general lines of the spectral transform and the Darboux transform were outlined; in this paragraph we want to briefly discuss an important example: the KdV.
First of all, it should be emphasized that the KdV is a model equation, together with the Hopf, Burgers equation and the NLS, of the non-linear mathematical physics; indeed these equations can be found by applying multiscale expansion to rather general classes of nonlinear PDEs; with reference to \cite{4}, the case of the KdV is reported in appendix \ref{appA}. 
\subsubsection{Soliton $N$ solution with the spectral transform for the KdV}
Let us consider the case of the KdV, which is generally written as
\begin{equation}
    u_{,t}-6uu_{,z}+u_{,z,z,z}=0
\label{KdV1}    
\end{equation}
with $u$ a function of $z$ and $t$, and we apply the TS to derive the one soliton solution. The Lax pair for the KdV can be written as
\begin{equation}
\begin{aligned}
    &\hat{L}\psi=(-\partial_z^2+u)\psi=E\psi,\\
    &\psi_{,t}=\hat{M}\psi=(-4\partial_z^3+6u\partial_z+3u_{,z}+c(E))\psi,
\label{LAXKdV}    
\end{aligned}
\end{equation}
where $c(E)_{,z}=0$; the integrability condition for the \ref{LAXKdV} problem returns exactly the \ref{KdV1} and moreover it is verified if and only if there is the isospectrality of the problem, that is $E_{,t}=0$:
\begin{equation}
\begin{aligned}
    &(\hat{L}\psi)_{,t}=\hat{L}_{,t}\psi+\hat{L}\psi_{,t}=(\hat{L}_{,t}+\hat{L}\hat{M})\psi=(E\psi)_{,t}=E_{,t}\psi+E\psi_{,t}=\\
    &=E_{,t}\psi+E\hat{M}\psi=(E_{,t}+\hat{M}\hat{L})\psi \Rightarrow E_{,t}+\hat{M}\hat{L}=\hat{L}_{,t}+\hat{L}\hat{M} \Rightarrow\\
    & \Rightarrow E_{,t}=0 \Leftrightarrow \hat{L}_{,t}+[\hat{L},\hat{M}]=0.
\end{aligned}    
\end{equation}
As can be seen from the \ref{LAXKdV} the spectral problem is that associated with the Schroedinger equation; the direct problem therefore consists in associating the spectral parameters given by the Schroedinger equation with potential $u(z,0)$. The evolution of the spectral data is always obtained from the \ref{LAXKdV} system: taking the derivative with respect to $z$ of the first equation we have
\begin{equation*}
-\partial_z^3\psi+u_{,z}\psi+u\psi_{,z}=E\psi_{,z} \Rightarrow  \psi_{,z,z,z}=u_{,z}\psi+(u-E)\psi_{,z},
\end{equation*}
which substituted in the second equation returns
\begin{equation}
\begin{aligned}
    &\psi_{,t}=-4u_{,z}\psi-4(u-E)\psi_{,z}+6u\psi_{,z}+3u_{,z}\psi+c(E)\psi \Rightarrow\\
    & \Rightarrow \psi_{,t}=(-u_{,z}+c(E))\psi+(2u+4E)\psi_{,z}.
\end{aligned}    
\end{equation}
The inverse problem instead passes through the analytic properties of the eigenfunctions. We define $E=k^2$; for the discrete spectrum we must have $E_n<0 $ and therefore $k_n=ip_n$. The inverse problem gives us the solution to $N$ solitons, which is obtained from the algebraic system of $N$ equations
\begin{equation}
(1+e^{-2p_n(z-4p_n^2t-\gamma_n)})f_n(z,t)+\sum_{m=1,\neq n}^N\frac{2p_me^{-2p_m(z-4p_m^2t-\gamma_m)}}{p_m+p_n}f_m(x,t)=1, n=1,...,N;
\label{SISS}
\end{equation}
where the $f_n(x,t)$ are the unknowns and the $\gamma_n$ of the constants that depend on the spectral parameters a $t=0$, together with the
\begin{equation}
    u(z,t)=4\partial_z\bigg(\sum_{n=1}^Np_ne^{-2p_n(z-4p_n^2-\gamma_n)}f_n(x,t)\bigg).
\label{uKdV}    
\end{equation}
The solution of the \ref{SISS} system can be found with Cramer's method and the \ref{uKdV} gives us the final result in the form
\begin{equation}
    u(z,t)=-2\partial_z^2ln(det_N),
\label{NSol}
\end{equation}
where $det_N$ is the determinant of the matrix associated with the $N \times N$  system \ref{SISS}).
For example the one soliton solution\footnote {It is interesting to note that the one soliton solution can be obtained directly by quadrature from the KdV.} is given by
\begin{equation}
    u(z,t)=-\frac{2p_1^2}{cosh^2(p_1(z-4p_1^2t-\gamma_1))}=2E_1sech^2(\sqrt{|E_1|}(z+4E_1t-\gamma_1));
\label{A}    
\end{equation}
the solution is therefore exponentially localized with amplitude and velocity proportional to $p_1^2$ and therefore to the eigenvalue $E_1$.

Starting from the \ref{NSol} it is possible to show that, in the interaction between $N$ solitons, the $n$-th soliton does not change its shape but experiences a shift equal to
\begin{equation}
    \delta x_n=-\frac{1}{p_n}\bigg\{\prod_{j=1}^{n-1} ln\bigg(\bigg|\frac{p_n+p_j}{p_n-p_j}\bigg|\bigg)-\prod_{j=n+1}^N ln\bigg(\bigg|\frac{p_n+p_j}{p_n-p_j}\bigg|\bigg) \bigg\}.
\end{equation}

\subsubsection{Solitonic solutions of KdV with Darboux transform}
In this section we will show, in the specific case of KdV, that the poles of the transmission coefficient are the discrete elements of the spectrum of the associated spectral problem and that the TD acts by adding poles to the transmission coefficient and therefore solitons to the background solution. \\
We are looking for a TD of the form
\begin{equation}
    \boldsymbol{\psi}=\hat{\chi}\boldsymbol{\psi}_b,
\label{TDKdV}    
\end{equation}
con $\boldsymbol{\psi}=\binom{\psi}{\psi_{,z}}$, $\boldsymbol{\psi}_b=\binom{\psi_b}{\psi_{b,z}}$ e $\hat{\chi}=\begin{pmatrix}
A & B  \\
C & D 
\end{pmatrix}$. \\
By making explicit the first equation of the \ref{TDKdV} system, deriving it twice with respect to $z$, imposing that $A$ and $B$ do not depend on $k^2$, passing through the Riccati equation reduced to Schroedinger equation from the transformation $A=-\frac{\phi_{,z}}{\phi}$, we obtain the Darboux transformation for the KdV
\begin{equation}
    u(z,t)=u_b(z,t)-2\bigg(\frac{\phi_{,z}}{\phi}\bigg)_{,z} \ , \ \psi(z,k,t)=\psi_{b,z}(z,k,t)-\frac{\phi_{,z}}{\phi}\psi_{b}(z,k,t),
\label{TDkdv}    
\end{equation}
where $\phi$ is the solution of the Schroedinger equation $\phi_{,z,z}=(u_b-k^2)\psi $. It can be shown that by choosing the null background solution and taking the solution $\phi=c_1e^{p(z-4p^2t)}+c_2^{-p(z-4p^2t)}$ for $k=ip$, with the right choice of the constants $c_1$ and $c_2$ we get back the solution \ref{A}.
We consider a generic background solution $u_b(z,t)$ and choose as $\psi_b(z,t)$ the Jost eigenfunction defined as
\begin{equation}
    \begin{cases}
  J_b(z,k,t)\simeq e^{-ikz} & per \ z \simeq -\infty \\
     J_b(z,k,t)\simeq a_b(k)e^{-ikz}+b_b(k)e^{ikz}  & per \ z \simeq +\infty 
\end{cases},
\label{J}
\end{equation}
and let $\phi$ be the corresponding solution of the Lax pair for $k=ip$ with the property that
\begin{equation*}
    \phi \simeq \alpha_{\pm}e^{-pz}+\beta_{\pm}e^{pz}, \ \ per \ z \simeq \pm \infty, 
\end{equation*}
then it follows, again for $z\simeq \pm \infty$,
\begin{equation}
    \frac{\phi_{,z}}{\phi}=\frac{-p\alpha_{\pm}e^{-pz}+p\beta_{\pm}e^{pz}}{\alpha_{\pm}e^{-pz}+\beta_{\pm}e^{pz}} \simeq \pm p.
\label{Q}    
\end{equation}
Using the \ref{J} and \ref{Q} in the second of \ref{TDkdv} you get a $z\simeq-\infty$
\begin{equation}
    \psi(z,k,t)\simeq -ike^{-ikz}-pe^{-ikz}=-i(k+ip)e^{-ikz}.
\label{psi}    
\end{equation}
Let $J(z,k,t)$ be the Jost eigenfunction corresponding to the potential $u(z,t)$, defined analogously to $J_b(z,k,t)$, that is
\begin{equation}
    \begin{cases}
  J(z,k,t)\simeq e^{-ikz} & per \ z \simeq -\infty \\
     J(z,k,t)\simeq a(k)e^{-ikz}+b(k)e^{ikz}  & per \ z \simeq +\infty 
\end{cases};
\label{J2}
\end{equation}
from \ref{psi}, \ref{J2}, from the second of \ref{TDkdv} and from \ref{J} we have
\begin{equation}
    \psi(z,k,t)\simeq -i(k+ip)J(z,k,t) \Rightarrow J(z,k,t)=\frac{1}{-i(k+ip)}\bigg(J_{b,z}(z,k,t)-\frac{\phi_{,z}}{\phi}J_{b}(z,k,t)\bigg)
\end{equation}
and from this, thanks to the \ref{J} and \ref{Q}, it immediately follows that for $z \simeq \infty $
\begin{equation}
\begin{aligned}
    &J(z,k,t) \simeq \frac{i}{k+ip}\bigg(-ik\big(a_b(k)e^{-ikz}-b_b(k)e^{ikz}\big)-p\big(a_b(k)e^{-ikz}+b_b(k)e^{ikz}\big)\bigg)=\\
    &=\frac{i}{k+ip}\bigg(a_b(k)e^{-ikz}\big(-ik-p\big)+b_b(k)e^{ikz}\big(ik-p\big)\bigg)=\\
    &=\frac{k-ip}{k+ip}a_b(k)e^{-ikz}+\frac{-k-ip}{k+ip}b_b(k)e^{ikz}=\frac{k-ip}{k+ip}a_b(k)e^{-ikz}-b_b(k)e^{ikz},
\end{aligned}    
\end{equation}
which must ultimately be the same as \ref{J2}
\begin{equation}
    a(k)=\frac{k-ip}{k+ip}a_b(k) \ , \ b(k)=-b_b(k).
\end{equation}
Since in general, the $a(k)$ and the $b(k)$ are related to the transmission and reflection coefficients by the
\begin{equation}
    T(k)=\frac{1}{a(k)} \ , \ R(k)=\frac{b(k)}{a(k)},
\end{equation}
we conclude that
\begin{equation}
    T(k)=\frac{k+ip}{k-ip}T_b(k) \ , \ R(k)=-R_b(k);
\end{equation}
the Darboux transform adds a pole placed in $ k=ip$ to the transmission coefficient and therefore a zero to $a(k)$.
We therefore consider the zero $k=ip:=k_0$; since $|a(k)|^2-|b(k)|^2=1$ and $a(k)$ is analytic in the upper half plane, we have $Im(k_0)>0$ and therefore $p>0$. From the definition of the Jost function it follows\footnote {See, for example, the \ref{J2}.} that
\begin{equation}
    \begin{cases}
  J(z,k_0,t)\simeq e^{-ik_0z}=e^{pz} \rightarrow 0  & per \ z \rightarrow -\infty \\
     J(z,k_0,t)\simeq a(k_0)e^{-ik_0z}+b(k_0)e^{ik_0z}=b(k_0)e^{-pz} \rightarrow 0 & per \ z \rightarrow +\infty 
\end{cases};
\label{J3}
\end{equation}
that is, normalizable functions which therefore correspond to discrete eigenvalues: the poles of the transmission coefficient are discrete elements of the spectrum. \\
The two results obtained allow us to conclude that TD adds solitons to the background solution.

\section{Gravitational solitons}\label{ch3}
In this chapter we look at the soliton solutions of Einstein's equations in vacuum: the so-called gravitational solitons. The application and extension of the TS technique to general relativity begins at the end of the 1970s thanks to the works of Belinski, Zakharov and Maison \cite{3},\cite{9},\cite{10} and continues with the extension in the context of the Maxwell-Einstein equations by Alekseev \cite{1} in the early 1980s. \\
Gravitational solitons have some interesting properties, some of these will be discussed in paragraph \ref{sub3.2}.
In the last section, with reference \cite{11}, we will discuss how the Kerr black hole can be interpreted as a two soliton solution of the field equations in the presence of axial symmetry.

\subsection{Generalization of the spectral transform by gravitation theory and solution to $N$ solitons}\label{sub3.1}

Let us start by establishing the problem. Einstein field equations are written
\begin{equation}
     E _{\mu \nu}=\frac{8 \pi G}{c^4}T_{\mu \nu},
\label{EEE}     
\end{equation}
the tensor $E_ {\mu \nu}$ is called the Einstein tensor and is defined as $E_{\mu \nu}:=R_{\mu \nu} - \frac{1}{2} g_{\mu \nu}R$ where $R_{\mu \nu}$ is the Ricci tensor and $T_{\mu \nu}$ is the energy-momentum tensor of the source. \\
The trace of the Einstein tensor returns $E=-R$ and, thanks to \ref{EEE}, we have $R=-\frac{8 \pi G}{c^4}T$ and using the \ref{EEE} again 
\begin{equation}
     E _{\mu \nu}=R_{\mu \nu}+\frac{1}{2}g_{\mu \nu}\frac{8 \pi G}{c^4}T=\frac{8 \pi G}{c^4}T_{\mu \nu} \Rightarrow R_{\mu \nu}=\frac{8 \pi G}{c^4}\bigg(T_{\mu \nu}-\frac{1}{2}g_{\mu \nu}T \bigg);
\end{equation}
in vacuum the momentum energy tensor is identically zero and the field equations can be expressed as
\begin{equation}
    R_{\mu \nu}=0.
\label{EV}    
\end{equation}
They are a system of differential equations for the 10 independent components of the metric and represent our starting problem\footnote{Let us assume that we are in a four-dimensional space-time.}. From here on we will use the geometric units in which $c=G=1$.

The TS method, defined in its general lines in section \ref{ch2}, is applicable only when the unknown function depends on two variables; in general, however, the metric depends on the three spatial coordinates and the time coordinate. We therefore assume that space-time admits two Killing vector fields of type space that allow us to eliminate the dependence on two spatial variables\footnote{Given a certain vector field, $\vec{\xi}(x^{\mu})$, it will identify a symmetry if and only if it satisfies the Killing equation:
\begin{equation*}
    g_{\mu \nu , \rho} \xi^{\rho}+g_{\beta \nu}\xi^{\beta}_{,\mu}+g_{\mu \beta}\xi^{\beta}_{,\nu}=0.
\end{equation*}
Let us consider that we have a Killing field of type time, $\vec{\xi}$, then we could choose as the coordinate system one in which the basic vector of type time, $\vec{e}_{(0)}$, is at any point aligned with the $\vec{\xi}$ field. In this reference the components of the vector field are $\xi^{\alpha}=(\xi^{0},0,0,0)$ and if we parameterize the curve to which the vector field is associated in such a way that it results $\xi^0=1$ then the metric, thanks to the Killing equation, is independent of time and therefore stationary (all derivatives of $\xi^{\alpha}$ are zero and its only non-zero component is the temporal one). An analogous reasoning can be made in the case of a space-type or light-type Killing field.}: $x^1$ and $x^2$. With this choice, the metric depends only on $x^0=t$ and $x^3=z$. To simplify the problem further we refer to the gauge symmetry of general relativity: symmetry under diffeomorphisms. In four-dimensional space-time this symmetry allows us to impose 4 conditions on the metric without loss of generality: we will impose that
\begin{equation}
    g_{00}=-g_{33}\ , \ g_{03}=0\ , \ g_{0i}=0,
\end{equation}
where $i=1,2$. \\ 
Under these hypotheses, the space-time interval is written as
\begin{equation}
    ds^2=g_{\mu \nu}dx^{\mu}dx^{\nu}=g_{00}(dt^2-dz^2)+g_{ij}dx^idx^j+2g_{i3}dx^idz,
\label{MS}
\end{equation}
in which, from here to the end of paragraph \ref{sub3.2}, $i,j=1,2$ (later on we will introduce the index $ h $, which also runs on 1 and 2). It is not yet known how to apply the TS to the \ref{MS} metric: further simplification is needed. Since we have exhausted our freedom to manipulate the metric without loss of generality, we must impose as a physical condition that $g_{i3}=0 $; this allows us to write
\begin{equation}
    ds^2=g_{00}(dt^2-dz^2)+g_{ij}dx^idx^j.
\label{MV}    
\end{equation}
We note that if we had chosen the Killing fields, one of the space like and one of the time type, we would have obtained a metric similar to \ref{MV} but stationary, in which the independent variables would be both space like; this case will be considered in paragraph \ref{sub3.3} to derive the Kerr metric.

We call $\hat{g}$ the $2 \times2$ block $g_{ij}$ and set $det(\hat{g})=\alpha^2$; the equations \ref{EV} for the metric \ref{MV} are divided into two sets which, passing to the light cone coordinates introduced by $\zeta:=\frac{1}{2}(z+t)$ e $\eta:=\frac{1}{2}(z-t)$, take the form
\begin{equation}
\begin{aligned}
    &\big(\alpha \hat{g}_{, \zeta}\hat{g}^{-1}\big)_{, \eta}+ \big(\alpha \hat{g}_{, \eta}\hat{g}^{-1}\big)_{, \zeta}=0; \\
    & (ln(g_{00}))_{,\zeta}(ln(\alpha))_{,\zeta}=(ln(\alpha))_{,\zeta,\zeta}+\frac{A^2}{4\alpha^2} \ , \ (ln(g_{00}))_{,\eta}(ln(\alpha))_{,\eta}=(ln(\alpha))_{,\eta,\eta}+\frac{B^2}{4\alpha^2},
\label{EQgeg00}    
\end{aligned}    
\end{equation}
where $\hat{A}:=-\alpha \hat{g}_{,\zeta} \hat{g}^{-1}$ and $\hat{B}:=\alpha \hat{g}_{,\eta} \hat{g}^{-1}$.
The first of the equations \ref{EQgeg00} gives us the block $2 \times 2$ and the second ones gives us $g_{00}$, once given $\hat{g}$. We also note that taking the trace of the first equation and making use of the fact that $det(\hat{g})=\alpha^2$ yields
\begin{equation}
    \alpha_{,\zeta,\eta}=0;
\label{detOnde}    
\end{equation}
the root of the determinant satisfies the wave equation\footnote{Given the definitions $\zeta:=\frac{1}{2}(z+t)$ and $\eta:=\frac{1}{2}(z-t)$ we have $\alpha_{,\zeta,\eta}=\frac{1}{4}[(\partial_z+\partial_t)(\partial_z-\partial_t)]\alpha=\frac{1}{4}(\partial^2_z-\partial^2_t)\alpha=0\Rightarrow(\partial^2_z-\partial^2_t)\alpha=0$.}. \\
The first of the \ref{EQgeg00} is a second order equation and is therefore equivalent to a system of two first order equations. These two equations can be obtained directly from the definitions of $\hat{A}$ and $\hat{B}$: for the former it is sufficient to replace the definitions in the first of the \ref{EQgeg00}; the second is instead the condition of integrability between $\hat{A}$ and $\hat{B}$ with respect to $\hat{g}$, that is
\begin{equation}
\begin{aligned}
    &\hat{B}_{,\zeta}-\hat{A}_{,\eta}=0; \\
    &\hat{A}_{,\eta}+\hat{B}_{,\zeta}+\alpha^{-1}[\hat{A},\hat{B}]-\alpha_{,\eta}\alpha^{-1}\hat{A}-\alpha_{,\zeta}\alpha^{-1}\hat{B}=0.
\label{EF}    
\end{aligned}    
\end{equation}
The \ref{EF} are just a rewrite of the \ref{EV} field equations.

The next step is the search for an appropriate Lax pair that defines the spectral problem and therefore depends on a complex parameter: the spectral parameter $\lambda$. For this purpose we define the operators
\begin{equation}
    \hat{D}_-:=\partial_{\zeta}-\frac{2\alpha_{,\zeta}\lambda}{\lambda-\alpha}\partial_{\lambda} \ , \ \hat{D}_+:=\partial_{\eta}+\frac{2\alpha_{,\eta}\lambda}{\lambda+\alpha}\partial_{\lambda}
\end{equation}
and consider the following system of equations
\begin{equation}
\begin{aligned}
    &\hat{D}_-\boldsymbol{\psi}=\frac{\hat{A}}{\lambda-\alpha}\boldsymbol{\psi}, \\
    &\hat{D}_+\boldsymbol{\psi}=\frac{\hat{B}}{\lambda+\alpha}\boldsymbol{\psi},
\label{SDP}    
\end{aligned}
\end{equation}
where $ \boldsymbol{\psi} =\boldsymbol{\psi}(\zeta,\eta,\lambda)$ is called generating matrix; for $\lambda=0$ the system reduces to
\begin{equation}
\begin{aligned}
    &\boldsymbol{\psi}_{,\zeta}(\zeta,\eta,0)=-\frac{\hat{A}}{\alpha}\boldsymbol{\psi}(\zeta,\eta,0) \Rightarrow \hat{A}=-\alpha \boldsymbol{\psi}_{,\zeta}(\zeta,\eta,0)\boldsymbol{\psi}^{-1}(\zeta,\eta,0); \\
    &\boldsymbol{\psi}_{,\eta}(\zeta,\eta,0)=\frac{\hat{B}}{\alpha}\boldsymbol{\psi}(\zeta,\eta,0) \Rightarrow \hat{B}=\alpha \boldsymbol{\psi}_{,\eta}(\zeta,\eta,0)\boldsymbol{\psi}^{-1}(\zeta,\eta,0)
\end{aligned}    
\end{equation}
and, compared with the definitions of $\hat{A}$ and $\hat{B}$, it allows us to conclude that the block $2 \times2 $ is nothing more than the generating matrix evaluated for $\lambda=0$:
\begin{equation}
    \hat{g}(\zeta,\eta)=\boldsymbol{\psi}(\zeta,\eta,0).
\label{D}    
\end{equation}

We know that for the integration procedure it is necessary to know a backgound solution\footnote{Solution of the first of the equations \ref{EQgeg00}.} $\Hat{g}_b$ from which, thanks to the definitions of $\hat{A}$ and $\hat{B}$ we get the $ \hat{A}_b$ and $\hat{B}_b$ matrices which we replace in the system \ref{SDP} allow us to get the background generating matrix $ \boldsymbol{\psi}_b$. We therefore operate a Darboux transform of the form
\begin{equation}
    \boldsymbol{\psi}=\hat{\chi}\boldsymbol{\psi}_b
\label{E}    
\end{equation}
which replaced in the system \ref{SDP} returns a system of equations for the dressing matrix
\begin{equation}
\begin{aligned}
&\hat{D}_-\hat{\chi}=\frac{(\hat{A}\hat{\chi}-\hat{\chi}\hat{A_b})}{\lambda-\alpha};\\
&\hat{D}_+\hat{\chi}=\frac{(\hat{B}\hat{\chi}-\hat{\chi}\hat{B_b})}{\lambda+\alpha}.
\label{SisD}
\end{aligned}    
\end{equation}
Putting together \ref{D} and \ref{E} we see that
\begin{equation}
    \hat{g}(\zeta,\eta)=\boldsymbol{\psi}(\zeta,\eta,0)=\hat{\chi}(\zeta,\eta,0)\boldsymbol{\psi}_b(\zeta,\eta,0)=\hat{\chi}(\zeta,\eta,0)\hat{g}_b(\zeta,\eta);
\label{Solf}    
\end{equation}
at this point it is important to remember that the metric is real and symmetric; this implies having to impose conditions for the dressing matrix\footnote{Normalization is chosen for the matrix such that $\hat{\chi}(\zeta,\eta,\lambda)=\mathcal{I}$ when $|\lambda| \rightarrow  \infty$.} and the generating matrix in order to respect the properties of the metric; these conditions are expressed as
\begin{equation}
    \hat{\chi}^*(\zeta,\eta,\lambda^*)=\hat{\chi}(\zeta,\eta,\lambda) \ , \ \boldsymbol{\psi}^*(\zeta,\eta,\lambda^*)=\boldsymbol{\psi}(\zeta,\eta,\lambda) \ , \ \hat{g}=\hat{\chi}(\zeta, \eta,\lambda)\hat{g}_b\hat{\chi}^{T}(\zeta,\eta,\alpha^2 /\lambda);
\label{RSC}    
\end{equation}
where $*$ indicate complex conjugation. We underline that, for the moment, condition $det(\hat{g})=\alpha^2$ has not yet been imposed; the condition is imposed by requiring that
\begin{equation}
    \hat{g}^{(f)}=\alpha \frac{\hat{g}}{\sqrt{det(\hat{g})}},
\label{gfis}    
\end{equation}
where the superscript $(f)$ underlines that this is the solution that respects all the physical properties of the metric: this is physical solution. Obviously, \ref{gfis} involves a transformation of the $\hat{A}$ and $\hat{B}$ matrices which results in new equations for the $g_{00}$ component which will return the physical component, $g_{00}^{(f)}$. \\

We then proceed by discussing the construction of the $N$ solitons solution. This solution emerges by imposing a rational dependence of the dressing matrix on the spectral parameter with a finite number of simple poles
\begin{equation}
    \hat{\chi}=\mathcal{I}+\sum_{k=1}^N\frac{\hat{O}_k}{\lambda-\lambda_k},
\label{Dress}
\end{equation}
where $\hat{O}_k$ does not depend on the spectral parameter; from the reality condition of the matrix $\hat{g}$, and therefore from the first of the \ref{RSC}, it follows that the poles are either real or appear as pairs of conjugate complexes. The final solution is given by the \ref{Solf} so
\begin{equation}
    \hat{g}(\zeta,\eta)=\bigg(\mathcal{I}-\sum_{k=1}^N\frac{\hat{O}_k}{\lambda_k}\bigg)\hat{g}_b(\zeta,\eta);
\label{K}    
\end{equation}
it remains therefore to determine the matrix $\hat{O}_k$. For this purpose, we plug \ref{Dress} in system \ref{SisD} and the equations must be satisfied at the pole; these equations gives us important information. \\
The first is obtained by noting that the operators $\hat{D _-}$ and $\hat{D_+}$ contain derivations with respect to $\zeta$, $\eta$ and $\lambda$ and that, when applied to the non-trivial part of the dressing matrix, they produce terms containing double poles, terms that are not present in the second member: it is necessary that the coefficients of the double poles cancel out. In fact, studying only the $k$-th term for linearity we have\footnote{Remember that $\hat{O}_k$ does not depend on $\lambda$.}
\begin{equation}
\begin{aligned}
    &\bigg(\partial_{\zeta}-\frac{2\alpha_{,\zeta}\lambda}{\lambda-\alpha}\partial_{\lambda}\bigg)\frac{\hat{O}_k}{\lambda-\lambda_k}=\frac{\hat{O}_{k,\zeta}}{\lambda-\lambda_k}+ \hat{O}_k\bigg(\frac{-\lambda_{,\zeta}}{(\lambda-\lambda_k)^2}\bigg)-\frac{2\alpha_{,\zeta}\lambda}{\lambda-\alpha}\frac{\hat{O}_k}{(\lambda-\lambda_k)^2};\\
    &\bigg(\partial_{\eta}+\frac{2\alpha_{,\eta}\lambda}{\lambda+\alpha}\partial_{\lambda}\bigg)\frac{\hat{O}_k}{\lambda-\lambda_k}=\frac{\hat{O}_{k,\eta}}{\lambda-\lambda_k}+ \hat{O}_k\bigg(\frac{-\lambda_{,\eta}}{(\lambda-\lambda_k)^2}\bigg)+\frac{2\alpha_{,\eta}\lambda}{\lambda+\alpha}\frac{\hat{O}_k}{(\lambda-\lambda_k)^2};
\end{aligned}    
\end{equation}
eliminating coefficients of the double poles calculated on the poles we have
\begin{equation}
\begin{aligned}
  &\bigg(-\lambda_{,\zeta}-\frac{2\alpha_{,\zeta}\lambda}{\lambda-\alpha}\bigg)\bigg|_{\lambda=\lambda_k}=0 \Rightarrow \lambda_{k,\zeta}=\frac{2\alpha_{,\zeta}\lambda_k}{\alpha-\lambda_k};\\  
  &\bigg(-\lambda_{,\eta}+\frac{2\alpha_{,\eta}\lambda}{\lambda+\alpha}\bigg)\bigg|_{\lambda=\lambda_k}=0 \Rightarrow \lambda_{k,
  \eta}=\frac{2\alpha_{,\eta}\lambda_k}{\alpha+\lambda_k}; 
\label{POLIm}  
\end{aligned}
\end{equation}
the \ref{POLIm} determine the trajectories of the poles: the poles are not fixed but move in space-time. Moreover, noting that $ \hat{O}_k$ and $\hat{\chi}^{- 1}(\lambda_k)$ are degenerate matrices\footnote{This conclusion can be seen starting from the relation $\hat{\chi}\hat{\chi}^{- 1}=\mathcal{I}$. In fact the identity matrix has no poles while the left side gives rise to the term $\sum_{k=1}^N\frac{\hat{O}_k \hat{\chi}^{-1}}{\lambda-\lambda_k}$ which contains simple poles: the residue must be null and therefore $\hat{O}_k\hat{\chi}^{-1}(\lambda_k)=0$, from which degeneration follows.} , then we can write them as
\begin{equation}
    (\hat{O}_k)_{ij}=n^{(k)}_im^{(k)}_j\ , \ (\hat{\chi}^{-1}_k(\lambda_k))_{ij}=q^{(k)}_ip^{(k)}_j;
\label{mdeg}    
\end{equation}
we rewrite the equations \ref{SisD} by multiplying them from the right by $\hat{\chi}^{-1}$
\begin{equation}
\begin{aligned}
&\frac{\hat{A}}{\lambda-\alpha}=(\hat{D}_-\hat{\chi})\hat{\chi}^{-1}+\hat{\chi}\frac{\hat{A}_b}{\lambda-\alpha}\hat{\chi}^{-1},\\
&\frac{\hat{B}}{\lambda+\alpha}=(\hat{D}_+\hat{\chi})\hat{\chi}^{-1}+\hat{\chi}\frac{\hat{B}_b}{\lambda+\alpha}\hat{\chi}^{-1};
\label{EQ}
\end{aligned}
\end{equation}
since the left sides of the equations \ref{EQ} are regular to the poles $\lambda=\lambda_k$, it follows that the residues of the poles in the second members must be zero, which leads to the equations
\begin{equation}
\begin{aligned}
    &\hat{O}_{k,\zeta}\hat{\chi}^{-1}(\lambda_k)+\hat{O}_k\frac{\hat{A}_b}{\lambda_k-\alpha}\hat{\chi}^{-1}(\lambda_k)=0,\\
    &\hat{O}_{k,\eta}\hat{\chi}^{-1}(\lambda_k)+\hat{O}_k\frac{\hat{B}_b}{\lambda_k+\alpha}\hat{\chi}^{-1}(\lambda_k)=0.
\label{SISO}    
\end{aligned}    
\end{equation}
Inserting the \ref{mdeg} into the \ref{SISO} equations we obtain the equations that determine the vectors $m_i^{(k)}$:
\begin{equation}
\begin{aligned}
    &\bigg(m^{(k)}_{i,\zeta}+m_j^{(k)}\frac{(\hat{A}_b)_{ij}}{\lambda_k-\alpha}\bigg)q_i^{(k)}=0,\\
    &\bigg(m^{(k)}_{i,\eta}+m_j^{(k)}\frac{(\hat{B}_b)_{ij}}{\lambda_k+\alpha}\bigg)q_i^{(k)}=0,
\end{aligned}
\end{equation}
whose solution is given by
\begin{equation}
    m_i^{(k)}=m_{0j}^{(k)}(\boldsymbol{\psi}_b^{-1}(\lambda_k,\zeta,\eta))_{ji}
\label{Z}    
\end{equation}
where the $m_{0j}^{(k)}$ are arbitrary complex constant vectors. To determine the $n_i^{(k)}$ the third condition of \ref{RSC} is exploited, which allows, ultimately, to write the vectors $n_i^{(k)}$ as
\begin{equation}
    n_i^{(k)}=\sum_{l=1}^N\frac{1}{\lambda_l}(\hat{\Pi})_{kl}L_i^{(l)},
\label{S}    
\end{equation}
where $L_i^{(l)}:=m_j^{(l)}(\hat{g}_b)_{ji}$ and $(\hat{\Pi})_{kl}:=(\hat{\Gamma}^{-1})_{kl}$ with $(\hat{\Gamma})_{kl}:=-m_i^{(k)}m_j^{(l)}(\hat{g}_b)_{ij}(\alpha^2-\lambda_k\lambda_l)^{-1}$.

From \ref{K}, using the first of the \ref{mdeg} and \ref{S} gives\footnote{Remember that the metric is symmetric.}
\begin{equation}
\begin{aligned}
    &(\hat{g})_{ij}=(\hat{g}_b)_{ij}-\sum_{k=1}^N\frac{(\hat{O}_k)_{ih}}{\lambda_k}(\hat{g}_b)_{hj}=(\hat{g}_b)_{ij}-\sum_{k=1}^N\frac{1}{\lambda_k}n_i^{(k)}m_{h}^{(k)}(\hat{g}_b)_{hj}=\\
    &=(\hat{g}_b)_{ij}-\sum_{k=1}^N\sum_{l=1}^N\frac{1}{\lambda_k}\frac{1}{\lambda_l}(\hat{\Pi})_{kl}L_i^{(l)}\underbrace{m_h^{(k)}(\hat{g}_b)_{hj}}_{=L_j^{(k)}}=(\hat{g}_b)_{ij}-\sum_{k=1}^N\sum_{l=1}^N\frac{1}{\lambda_k}\frac{1}{\lambda_l}(\hat{\Pi})_{kl}L_i^{(l)}L_j^{(k)};
\label{jk}    
\end{aligned}    
\end{equation}
it should be stressed, however, that the solution \ref{jk} is not the physical one: condition $det(\hat{g})=\alpha^2$ is not satisfied. To obtain the physical solution we refer to \ref{gfis}; the computation of the determinant of $\hat{g}$ can be performed step by step: it is first computed for the one soliton solution, then the physical soliton solution is used as the background solution and the two soliton solution is constructed , the determinant is computed and the two soliton solution is used as the backgound solution to add another soliton, the determinant is computed and so on\footnote {The computation yields $det(\hat{g})=\alpha^{2N+2}\prod_{k=1}^N\frac{1}{\lambda_k^2}$.}.
As already mentioned, the calculation of the solution $\hat{g}^{(f)}$ implies that the coefficient $g_{00}$ must also be recalculated to obtain the physical coefficient $g_{00}^{(f) }$. In conclusion we get
\begin{equation}
\begin{aligned}
    &\hat{g}^{(f)}=\frac{\hat{g}}{\alpha^N}\prod_{k=1}^N\lambda_k; \\
    &g_{00}^{(f)}=Cg_{b00}\alpha^{-\frac{N^2}{2}}\bigg(\prod_{k=1}^N\lambda_k\bigg)^{N+1}\bigg(\prod_{k>l=1}^N\frac{1}{(\lambda_k-\lambda_l)^2}\bigg)det(\hat{\Gamma});
\label{SOLF}    
\end{aligned}    
\end{equation}
where $C$ is an arbitrary constant to choose with appropriate sign in order to have the correct sign for $g^{(f)}_{00}$; $g_{b00}$ is the $g_{00}$ background solution corresponding to $\hat{g}_b$; $\hat{g}$ is provided by \ref{jk}.

The solution to $N$ solitons is therefore given by replacing the \ref{SOLF} in the space-time interval \ref{MV} and remembering the \ref{jk}
\begin{equation}
\begin{aligned}
    &ds^2=g^{(f)}_{00}(dt^2-dz^2)+g^{(f)}_{ij}dx^idx^j=\\
    &=Cg_{b00}\alpha^{-\frac{N^2}{2}}\bigg(\prod_{k=1}^N\lambda_k\bigg)^{N+1}\bigg(\prod_{k>l=1}^N\frac{1}{(\lambda_k-\lambda_l)^2}\bigg)det(\hat{\Gamma})(dt^2-dz^2)+\\
    &+\frac{\big((\hat{g}_b)_{ij}-\sum_{k,l=1}^N \lambda_k^{-1}\lambda_l^{-1}(\hat{\Pi})_{kl}L_i^{(k)}L^{(l)}_j\big)}{\alpha^N}\bigg(\prod_{k=1}^N\lambda_k\bigg)dx^idx^j;
\end{aligned}    
\label{MVF}    
\end{equation}
where, $L_i^{(l)}:=m_j^{(l)}(\hat{g}_b)_{ji}$ and $(\hat{\Pi})_{kl}:=(\hat{\Gamma}^{-1})_{kl}$ with $(\hat{\Gamma})_{kl}:=-m_i^{(k)}m_j^{(l)}(\hat{g}_b)_{ij}(\alpha^2-\lambda_k\lambda_l)^{-1}$.
\subsection{Some properties of gravitational solitons}\label{sub3.2}

Given built the $N$ solitons solution, we now want to discuss some interesting properties of gravitational solitons. \\
We explicitly report the solution to $N=1$ solitons; with reference to the \ref{SOLF} we have, considering that in this case $\Pi$ and $\Gamma$ are not matrices:
\begin{equation}
\begin{aligned}
   &({\hat{g}}^{(f1)})_{ij}=\frac{(\hat{g}_b)_{ij}}{\alpha}\lambda_1-\bigg(\frac{(\alpha^2-\lambda_1^2)L_iL_j\lambda_1}{-\alpha \lambda_1^2m_im_j(\hat{g}_b)_{ij}}\bigg)=\frac{(\hat{g}_b)_{ij}}{\alpha}\lambda_1+\bigg(\frac{(\alpha^2-\lambda_1^2)L_iL_j}{\alpha \lambda_1m_im_j(\hat{g}_b)_{ij}}\bigg),\\
   &\hat{g}_{00}^{(f1)}=\Tilde{C}g_{b00}\alpha^{-\frac{1}{2}}\lambda_1^2m_im_j(\hat{g}_b)_{ij}(\alpha^2-\lambda_1^2)^{-1},
\label{SOL1S}  
\end{aligned}
\end{equation}
where the superscript $(f1)$ indicates the physical solution to a soliton. The trajectory of the pole $\lambda_1$ is given by the equations \ref{POLIm} with $k=1$; the solution of these equations is given by the roots of the quadratic equation\footnote{For simplicity it will be shown only in the case of axisymmetric stationary metrics in paragraph 3.3.}
\begin{equation}
    \lambda_1^2+2(\beta-w_1)\lambda_1+\alpha^2,
\label{eqq}    
\end{equation}
where $w_1$ is an arbitrary constant and $\alpha$ and $\beta$ are two independent solutions of the equation \ref{detOnde}.
The solutions of \ref{eqq} are
\begin{equation}
\begin{aligned}
&\lambda_1^{(in)}=-(\beta-w_1)+\sqrt{(\beta-w_1)^2-\alpha^2}=(w_1-\beta)\bigg(1-\sqrt{1-\frac{\alpha^2}{(\beta-w_1)^2}}\bigg)\\
&\lambda_1^{(out)}=-(\beta-w_1)-\sqrt{(\beta-w_1)^2-\alpha^2}=(w_1-\beta)\bigg(1+\sqrt{1-\frac{\alpha^2}{(\beta-w_1)^2}}\bigg),
\label{Solp}
\end{aligned}
\end{equation}
apexes $(in)$ and $(out)$ underline that the solutions are inside or outside the circle generated by $|\lambda|=\alpha$; specifically $|\lambda_1^{(in)}| < \alpha$ and $|\lambda_1^{(out)}| > \alpha$. In the figure below\footnote {The graph was created with GeoGebra.} we shows the case where $\lambda_1^{(in)}$ and $\lambda_1^{(out)}$ are real with $w_1=3$ and $\alpha=2$.
\begin{figure}[H]
    \centering
    \includegraphics[width = 15cm]{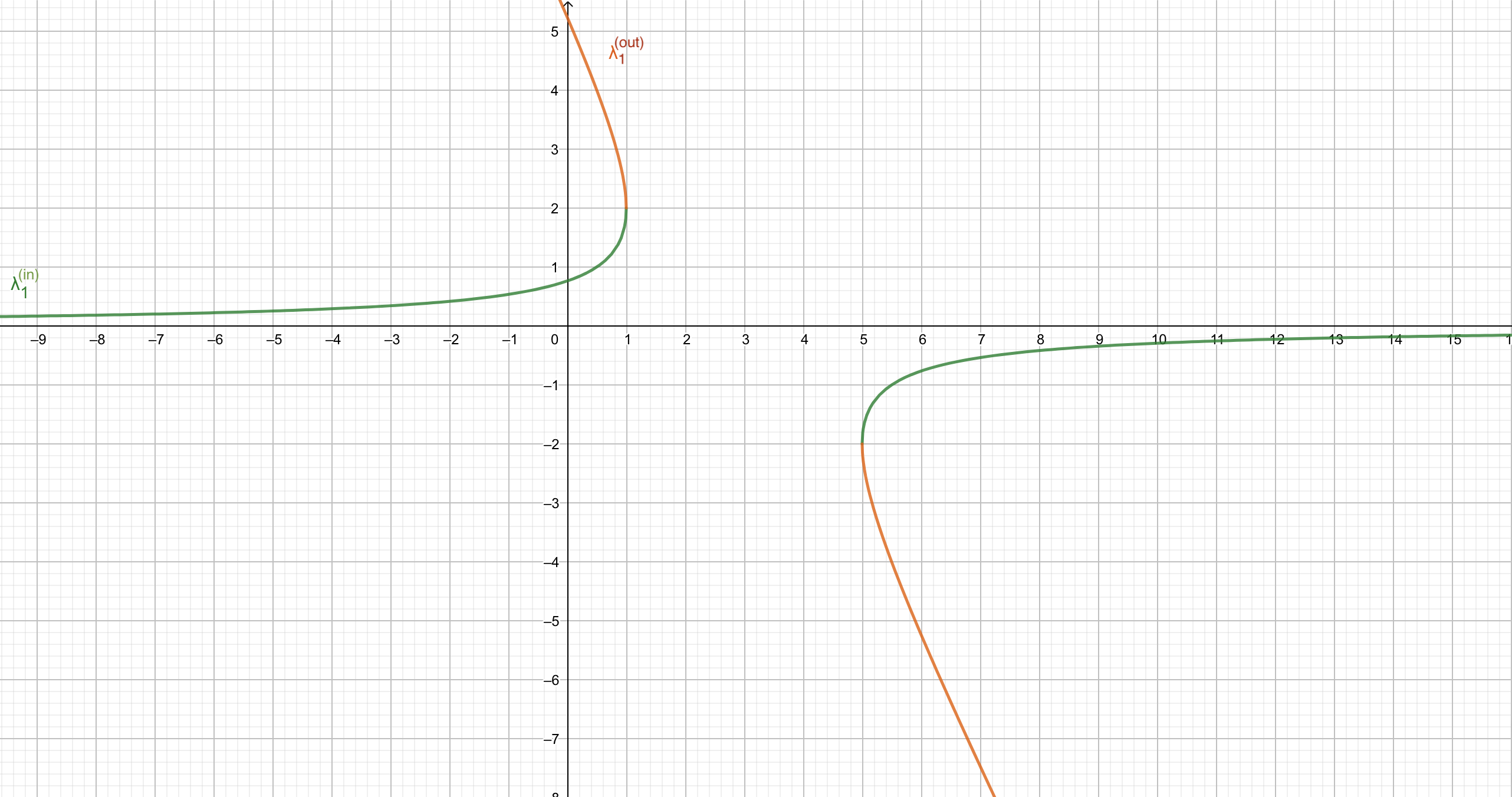}
    \caption{\textit{Andamento di $\lambda_1^{(in)}$ (linea verde) e $\lambda_1^{(out)}$ (linea arancione) in funzione di $\beta$ nel caso in cui siano reali, con $w_1=3$ e $\alpha=2$. Si nota che $|\lambda_1^{(in)}|$ è sempre minore di $\alpha=2$ mentre $|\lambda_1^{(out)}|$ è sempre maggiore di $\alpha=2$; in generale, $\lambda_1^{(in)}$ è sempre contenuto nella regione $|\lambda|=\alpha$ mentre  $\lambda_1^{(out)}$ è sempre esterno a tale regione}.}
    \label{GP}
\end{figure}
Let us consider the one soliton solution generated by a real pole: looking at the \ref{Solp} we see that the solutions, in the case of a real pole, are defined only in the region $(\beta-w_1)^2 \geq \alpha^2$, however the solution can be continued analytically in the $(\beta-w_1)^2 < \alpha^2$ region. When passing between the two regions, the $\lambda_1$ function becomes complex:
\begin{equation}
    \lambda_1=(w_1-\beta)\bigg(1\pm i \sqrt{\frac{\alpha^2}{(\beta-w_1)^2}-1}\bigg)
\end{equation}
and since the complex poles of the dressing matrix can appear only in pairs, it follows that, necessarily, in the region $(\beta-w_1)^2 < \alpha^2$ there is a solution a $N=2$ solitons generated by two complex poles conjugate $\lambda_1$ and $\lambda_1^*$. The modulus of these poles, however, is equal to $\alpha^2$, in fact
\begin{equation}
    |\lambda_1|=|\lambda_1^*|=(w_1-\beta)^2+(w_1-\beta)^2\bigg(\frac{\alpha^2}{(\beta-w_1)^2}-1\bigg)=\alpha^2,
\end{equation}
and from \ref{S}, remembering the definitions of $L_i^{(l)}$, $(\hat{\Pi})_{kl}$ e $(\hat{\Gamma})_{kl}$, it follows that $n_i^{(k)}$ are null if $|\lambda_k|=\alpha^2$ and therefore the dressing matrix is reduced to identity if the poles lie on the circumference generated by $|\lambda|=\alpha$. This means that the two soliton solution actually reduce to the background solution; we also note that when $(\beta-w_1)^2=\alpha^2$ we have $|\lambda_1|=\alpha^2$ and again the solution reduce to the background solution. From \ref{SOL1S} we see that the coefficient $g_{00}^{(f1)}$ is singular when $|\lambda_1|=\alpha^2$. \\
To recap: in the region $(\beta-w_1)^2 \geq \alpha^2$ we have the one soliton solution while in the region $(\beta-w_1)^2 < \alpha^2$ the solution is the background one; moreover in the region defined by the relation $(\beta-w_1)^2=\alpha^2$ the metric experiences a discontinuity due to the coefficient $g_{00}^{(f1)}$; in this sense the gravitational solitons can be considered as shock waves\footnote{Shock waves arise from the regularization of the solution of a certain model: a discontinuous but monodrome solution is preferred to a continuous solution but a polyhydrome, this is the shock wave.} gravitational. Similar considerations are possible for the case of $N$ solitons: there are regions in which we have the $N$ solitons solution, regions in which we have the $N-1$ solitons solution and so on; moreover, the metric is discontinuous in each interface between the various regions. However, it should be emphasized that this is true only for solutions generated by real poles; if the poles were all complexes conjugated in pairs there would be no regions of discontinuity and all space-time would be perturbed with respect to the background metric.

A second interesting property is the so-called fusion of poles. Let us consider the solution a $N=2$ solitons; suppose that the constants $m^{(1)}_{0i}$ and $m^{(2)}_{0i}$ depend, respectively, on $w_1$ and $w_2$ and that in the limit $w_2\rightarrow w_1$ we have $m^{(2)}_{0i} \rightarrow m^{(1)}_{0i}$, consequently we have $\lambda_2 \rightarrow \lambda_1$. It is possible to show from the solution of \ref{SOL1S} with $N=2$ that this does not vanish under the limit procedure just described but rather reduces to a one soliton solution corresponding to a double pole. This conclusion is interesting and is a non-trivial and unexpected phenomenon: we started by considering only simple poles and we find ourselves with the possibility of double poles thanks to the fusion of poles. The same analysis can be done in the case of the $N$ solitons solution and of the fusion of the $N$ poles which leads to a one soliton solution corresponding to a pole of order $N$. The fusion of the poles can occur in different ways, for example if the solution has $N=2n$ solitons and therefore $2n$ poles it is possible to fuse poles separately to form two poles of order $n$, and therefore two $n$ solitons solutions.

The last properties that we want to deal with, even if very briefly, are the topological properties of gravitational solitons. Specifically, we want to focus on the fact that it is not clear when a gravitational soliton is a topological object and a topological charge can be associated with it. In general, a topological soliton is a solution that presents itself with a different topology from the vacuum topology of the problem; to these objects it is possible to associate quantities, called topological charges\footnote {Topological charges, are quantities that characterize the topology of the soliton. In general, topological charges are topological invariants such as homotopy groups (defined as the quotient group between collapsed closed dimensional loops $d$ with the equivalence relationship according to which two loops are equivalent if and only if they are homotopes, i.e. transformable one into the other with continuity) or homology groups.}, which characterize the topology.

\subsection{Kerr black hole as double gravitational soliton}\label{sub3.3}
In this paragraph we will consider an axisymmetric stationary metric, which therefore admits two Killing fields: one is space like and one is time like. In this section we will use the notation $(x^0,x^1,x^2,x^3)=(t,\phi,\rho,z)$ for the coordinates. With considerations similar to those made in paragraph 3.1 we can write the metric in the form \footnote{We have set $g_{22}=g_{33}$,$g_{i2}=g_{i3}=g_{23}=0$.}
\begin{equation}
    ds^2=g_{00}(d\rho^2+dz^2)+g_{ij}dx^idx^j,
\label{MVAS}    
\end{equation}
where, from here to the end of the paragraph, $i,j=0,1$ and where $g_{00}$ and $g_{ij}$ depend only on $x^2=\rho$ and $x^3=z$ which are both space-type variables. For stationary metrics of the type \ref{MVAS} it can be imposed, without loss of generality, that the block $2 \times 2$ $(\hat{g})_{ij}$ has $det(\hat{g})=-\rho^2$. This is a slightly different choice compared to the case seen in paragraph \ref{sub3.1}, because in the case of non-stationary metrics the determinant of the $2 \times 2$ block could be space like or time like, while in the case of stationary metrics it can only be time like. \\
Einstein's equations in vacuum for the \ref{MVAS} metric can be rewritten as
\begin{equation}
\begin{aligned}
    &(\rho \hat{g}_{,\rho}\hat{g}^{-1})_{,\rho}+(\rho \hat{g}_{,z}\hat{g}^{-1})_{,z}=0,\\
    &(ln(g_{00}))_{,\rho}=-\frac{1}{\rho}+\frac{1}{4\rho}(U^2-V^2) \ , \ (ln(g_{00}))_{,z}=\frac{1}{2\rho}UV;
\label{EVAS}    
\end{aligned}    
\end{equation}
where $\hat{U}:=\rho \hat{g}_{,\rho} \hat{g}^{- 1}$ and $\hat{V}:=\rho \hat{g}_{,z}\hat{g}^{-1}$. With steps analogous to those described in paragraph \ref{sub3.1}, the $N$ solitons solution is obtained in the case of the metric $\ref{MVAS}$
\begin{equation}
    ds^2=g_{00}^{(f)}(d\rho^2+dz^2)+g_{ij}^{(f)}dx^idx^j
\label{VB}    
\end{equation}
with
\begin{equation}
\begin{aligned}
    &g_{00}^{(f)}=16Cg_{b00}\rho^{-\frac{N^2}{2}}\bigg(\prod_{k=1}^N\lambda_k\bigg)^{N+1}\bigg(\prod_{k>l=1}^N\frac{1}{(\lambda_k-\lambda_l)^2}\bigg)det(\hat{\Gamma});\\
    &(\hat{g})_{ij}^{(f)}=\pm \frac{\hat{g}_{ij}}{\rho^N}\bigg(\prod_{k=1}^N \lambda_k\bigg)=\pm \frac{\big((\hat{g}_b)_{ij}-\sum_{k,l=1}^N(\hat{D})_{kl}\lambda_k^{-1}\lambda_l^{-1}L_i^{(k)}L_j^{(l)}\big)}{\rho^N}\bigg(\prod_{k=1}^N \lambda_k\bigg),
\label{corfs}    
\end{aligned}    
\end{equation}
where $(\hat{D})_{kl}:=(\hat{\Gamma}^{-1})_{kl}$, $(\hat{\Gamma})_{kl}:=m^{(k)}_i(\hat{g}_b)_{ij}m_j^{(l)}(\rho^2+\lambda_k \lambda_l)^{-1}$, $L_i^{(k)}:=m_j
^{(k)}(\hat{g}_b)_{ji}$, $m_i^{(k)}=m_{0j}^{(k)}(\boldsymbol{\psi}_b^{-1}(\lambda_k,\rho,z))_{ji}$. Matrix $(\hat{g}_b)_{ij}$ is the background solution and $g_{b00}$ is the solution $g_{00}$ corresponding to $(\hat{g}_b)_{ij}$ given by \ref{EVAS} evaluated with $\hat{U}_b:=\rho \hat{g}_{b,\rho}\hat{g}_b^{-1}$ and $\hat{V}_b:=\rho \hat{g}_{b,z}\hat{g}_b^{-1}$; moreover $\boldsymbol{\psi}^{-1}(\lambda, \rho, z)$ is the inverse of the solution generating matrix of the system\footnote {To obtain the background generating matrix just replace the matrices $\hat{U}$ and $\hat{V}$ with their background counterpart, that is $\hat{U}_b:=\rho \hat{g}_{b,\rho}\hat{g}_b^{-1}$ and $\hat{V}_b:=\rho \hat{g}_{b,z}\hat{g}_b^{-1}$.}
\begin{equation}
\begin{aligned}
\bigg(\partial_z-\frac{2\lambda^2}{\lambda^2+\rho^2}\partial_{\lambda}\bigg)\boldsymbol{\psi}=\frac{\rho \hat{V}-\lambda \hat{U}}{\lambda^2+\rho^2}\boldsymbol{\psi} \ , \ \bigg(\partial_{\rho}+\frac{2\lambda \rho}{\lambda^2+\rho^2}\partial_{\lambda}\bigg)\boldsymbol{\psi}=\frac{\rho \hat{V}+\lambda \hat{U}}{\lambda^2+\rho^2}\boldsymbol{\psi}.   
\label{BNM}
\end{aligned}    
\end{equation}
Furthermore, $C$ is an arbitrary constant but with a sign such that we have $g_{00}^{(f)} \geq 0$; the $+$ or $-$ sign in front of the second of the \ref{corfs} must be chosen appropriately in order to have the correct signature of the metric. Solitons of the form \ref{VB} are called stationary solitons.

A first difference with respect to the general case treated in paragraph \ref{sub3.1} is given by the equations that describe the trajectories of the poles, which in the case in question are written
\begin{equation}
\begin{aligned}
&\lambda_{k,z}=\frac{-2\lambda_k^2}{\lambda_k^2+\rho^2},\\
&\lambda_{k,\rho}=\frac{2\lambda_k\rho}{\lambda_k^2+\rho^2};
\label{zzz}
\end{aligned}
\end{equation}
the solution of these equations is given by the solutions of the quadratic equation (verified in appendix \ref{B})
\begin{equation}
    \lambda_k^2+2(z-w_k))\lambda_k-\rho^2=0,
\label{EQ2}    
\end{equation}
where the $w_k$ are complex arbitrary constants. The solutions of the \ref{EQ2} equation are
\begin{equation}
    \lambda_k=(w_k-z)\pm \sqrt{(w_k-z)^2+\rho^2},
\label{zzzz}    
\end{equation}
as we can see the root is always positive for real poles ($w_k$ real) and therefore the solitonic solution, in this case, will not experience discontinuity and it will be present in all space-time, contrary to what we saw in paragraph \ref{sub3.2}.

The second difference derives from the requirement that the condition $det(\hat{g})=-\rho^2$, necessary to have a physical solution, be satisfied. As discussed in paragraph \ref{sub3.1}, to obtain the physical solution, $\hat{g}^{(f)}$, we need to compute the determinant of the solution $\hat{g}$; the calculation leads to
\begin{equation}
    det(\hat{g})=(-1)^N\rho^{2N}\bigg(\prod_{k=1}^N\frac{1}{\lambda_k^2}\bigg)det(\hat{g}_b).
\label{der}    
\end{equation}
The \ref{der} shows that if the background solution is such that $det(\hat{g}_b)=-\rho^2$ then the solution must necessarily be an even number of solitons, $N=2n$, otherwise the sign of the determinant of $\hat{g}$ would change and this would lead to a non-physical metric. \\

We now come to the Kerr black hole. As mentioned earlier, in the case of axisymmetric stationary metrics the simplest solution is the one at $N=2$ solitons. We choose as the background metric the flat metric in cylindrical coordinates
\begin{equation}
    ds^2=-dt^2+\rho^2d\phi^2+d\rho^2+dz^2,
\end{equation}
comparing with \ref{MVAS} we see that $g_{b00}=1$ and that $\hat{g}_b=diag(-1,\rho^2)$ (which trivially satisfies $det(\hat{g}_b)=-\rho^2$). The arrays $\hat{V}_b$ and $\hat{U}_b$ are given by
\begin{equation}
\begin{aligned}
    &\hat{V}_b:=\rho \hat{g}_{b,z}\hat{g}_b^{-1}=0,\\
    &\hat{U}_b:=\rho \hat{g}_{b,\rho}\hat{g}_b^{-1}=\rho \begin{pmatrix}
0 & 0 \\
0 & 2\rho 
\end{pmatrix}\begin{pmatrix}
-1 & 0 \\
0 & \frac{1}{\rho^2} 
\end{pmatrix}=\begin{pmatrix}
0 & 0 \\
0 & 2
\end{pmatrix};
\end{aligned}    
\end{equation}
that used in the \ref{BNM} allow us to calculate the background generating matrix using
\begin{equation}
\begin{aligned}
&\bigg(\partial_z-\frac{2\lambda^2}{\lambda^2+\rho^2}\partial_{\lambda}\bigg)\begin{pmatrix}
A & B \\
C & D 
\end{pmatrix}=\frac{-\lambda}{\lambda^2+\rho^2}\begin{pmatrix}
0 & 0 \\
0 & 2
\end{pmatrix}\begin{pmatrix}
A & B \\
C & D 
\end{pmatrix};\\
&\bigg(\partial_{\rho}+\frac{2\lambda \rho}{\lambda^2+\rho^2}\partial_{\lambda}\bigg)\begin{pmatrix}
A & B \\
C & D 
\end{pmatrix}=\frac{\lambda}{\lambda^2+\rho^2}\begin{pmatrix}
0 & 0 \\
0 & 2
\end{pmatrix}\begin{pmatrix}
A & B \\
C & D 
\end{pmatrix},
\end{aligned}
\end{equation}
in which it is written $\boldsymbol{\psi}_b=\begin{pmatrix}
A & B \\
C & D 
\end{pmatrix}$;
by making all the equations explicit, the solution is obtained
\begin{equation}
    \boldsymbol{\psi}_b=\begin{pmatrix}
-1 & 0 \\
0 & \rho^2-2z\lambda-\lambda^2 
\end{pmatrix};
\end{equation}
from the matrix $\boldsymbol{\psi}_b$, remembering the definitions of  $m_i^{(k)}$, of $L_i^{(k)}$ and of $(\hat{\Gamma})_{kl}$ and using \ref{EQ2}, which gives us $\lambda^2+2z \lambda-\rho^2=2w_k\lambda_k$, we get\footnote{Remember that $\hat{g}_b=diag(-1, \rho^2)$.}
\begin{equation}
\begin{aligned}
&m_0^{(k)}=m_{0j}^{(k)}(\boldsymbol{\psi}^{-1}_b)_{j0}=m_{00}^{(k)}(\boldsymbol{\psi}_b^{-1}(\lambda_k,\rho,z))_{00}+m_{01}^{(k)}(\boldsymbol{\psi}_b^{-1}(\lambda_k,\rho,z))_{01}=-m_{00}^{(k)}:=C_{0}^{(k)};\\
\\
&m_1^{(k)}=m_{0j}^{(k)}(\boldsymbol{\psi}^{-1}_b)_{j1}=m_{00}^{(k)}(\boldsymbol{\psi}_b^{-1}(\lambda_k,\rho,z))_{01}+m_{01}^{(k)}(\boldsymbol{\psi}_b^{-1}(\lambda_k,\rho,z))_{11}=\\
&=\frac{m_{01}^{(k)}}{\rho^2-2z\lambda_k-\lambda_k^2}=\frac{m_{01}^{(k)}}{-2w_k\lambda_k}:=C_1^{(k)}\lambda_k^{-1};\\
\\
&L_0^{(k)}:=m_j^{(k)}(\hat{g}_b)_{j0}=m_0^{(k)}(\hat{g}_b)_{00}+m_1^{(k)}(\hat{g}_b)_{10}=-m_0^{(k)}=-C_0^{(k)};\\
\\
&L_1^{(k)}:=m_j^{(k)}(\hat{g}_b)_{j1}=m_0^{(k)}(\hat{g}_b)_{01}+m_1^{(k)}(\hat{g}_b)_{11}=m_1^{(k)}\rho^2=C_1^{(k)}\lambda_k^{-1}\rho^2;\\
\\
&(\hat{\Gamma})_{kl}:=m^{(k)}_i(\hat{g}_b)_{ij}m_j^{(l)}(\rho^2+\lambda_k \lambda_l)^{-1}=\\
&=m^{(k)}_0(\hat{g}_b)_{00}m_0^{(l)}(\rho^2+\lambda_k \lambda_l)^{-1}+m^{(k)}_1(\hat{g}_b)_{11}m_1^{(l)}(\rho^2+\lambda_k\lambda_l)^{-1}=\\
&=(\rho^2+\lambda_k\lambda_l)^{-1}(-C_0^{(k)}C_0^{(l)}+C_1^{(k)}\lambda_k^{-1}C_1^{(l)}\lambda_l^{-1}\rho^2).
\label{all}
\end{aligned}    
\end{equation}
The \ref{all} are all we need together with the \ref{EQ2}, for the pole values, and the \ref{corfs}, for the physical quantities, to build the $N=2n$ stationary solitons solution  with flat background metric. \\

Let us consider the case $N=2$ and write the two constants $w_1$ and $w_2$ of the two poles as
\begin{equation}
    w_1=\sigma_1+\sigma_2 \ , \ w_2=\sigma_1-\sigma_2;
\label{pop}    
\end{equation}
$\sigma_1$ is always taken real while $\sigma_2$ can be real (the poles will be real) or imaginary (the poles will be complex conjugated). We introduce two new variables $\theta$ and $r$ defined by the relations\footnote {The root is defined in such a way that in the limit $r\rightarrow \infty$ the dominant term returns the usual transformations between spherical and cylindrical coordinates.}
\begin{equation}
    \rho=sin(\theta)\sqrt{(r-m)^2-\sigma_2^2} \ , \ z=\sigma_1+(r-m)cos(\theta),
\label{rhozeta}
\end{equation}
where $m$ is an arbitrary constant. By inserting the \ref{pop} in the \ref{EQ2}, using the \ref{rhozeta}, we obtain
\begin{equation}
\begin{aligned}
&\lambda_1=2(r-m+\sigma_2)sin^2(\theta/2),\\
&\lambda_2=2(r-m-\sigma_2)sin^2(\theta/2);
\end{aligned}    
\end{equation}
in which the same sign has been chosen for the two poles.
Without loss of generality we can fixed
\begin{equation}
    C_1^{(1)}C_0^{(2)}-C_0^{(1)}C_1^{(2)}=\sigma_2 \ , \  C_1^{(1)}C_0^{(2)}+C_0^{(1)}C_1^{(2)}=-m,
\label{LO}
\end{equation}
and define the new arbitrary constants
\begin{equation}
    -b:=C_1^{(1)}C_1^{(2)}-C_0^{(1)}C_0^{(2)} \ , \  a:=C_1^{(1)}C_1^{(2)}+C_0^{(1)}C_0^{(2)};
\label{SO}     
\end{equation}
where $C_0^{(1)}$, $C_0^{(2)}$, $C_1^{(1)}$ and $C_1^{(2)}$ are the constants that appear in the first two equations some \ref{all};
putting together \ref{LO} and \ref{SO} we get
\begin{equation}
    \sigma_2^2=m^2-a^2+b^2.
\label{Oriz}    
\end{equation}
From \ref{corfs} thanks to \ref{all} we can calculate the physical quantities $(\hat{g})_{ij}^{(f)}$ and $g_{00}^{(f)}$ which, substituted in the line element \ref{MVAS} expressed in the coordinates $(t,\phi,\theta,r)$, allow to obtain
\begin{equation}
\begin{aligned}
    &ds^2=-C\omega\Delta^{-1}dr^2-C\omega d\theta^2-\omega^{-1}[\Delta-a^2sin^2(\theta)][dt+2ad\phi]^2+\\
    &+\omega^{-1}[4\Delta bcos(\theta)-4asin^2(\theta)[mr+b^2]][dt+2ad\phi]d\phi+\\
    &-\omega^{-1}[\Delta[asin^2(\theta)+2bcos(\theta)]^2-sin^2(\theta)[r^2+b^2+a^2]^2]d\phi^2,
\label{mnb}    
\end{aligned}
\end{equation}
in which the quantities $\omega:=r^2+[b-acos(\theta)]^2$ and $\Delta:=r^2-2mr+a^2-b^2 $ are defined. Setting $b=0$, redefining the time coordinate as $t:=t+2a\phi$ and setting $C=-1$ we have
\begin{equation}
\begin{aligned}
    &ds^2=\omega\Delta^{-1}dr^2+\omega d\theta^2-\omega^{-1}[\Delta-a^2sin^2(\theta)]dt^2-4mra\omega^{-1}sin^2(\theta)dtd\phi+\\
    &-\omega^{-1}[\Delta a^2sin^4(\theta)-sin^2(\theta)[r^2+a^2]^2]d\phi^2,
\label{mnb0}    
\end{aligned}
\end{equation}
where, now, $\omega=r^2+a^2cos^2(\theta)$ e $\Delta=r^2-2mr+a^2$. For convenience we rewrite \ref{mnb0} as
\begin{equation}
    ds^2=-dt^2+\omega\bigg(\frac{1}{\Delta}dr^2+d\theta^2\bigg)+(r^2+a^2)sin^2(\theta)d\phi^2+\frac{2mr}{\omega}[asin^2(\theta)d\phi-dt]^2;
\label{Kerr}    
\end{equation}
we immediately notice the presence of two possible singularities placed in $\Delta=0$ and $\omega=0$. The singularity in $\omega=0$ is a gravitational singularity\footnote{The singularity is placed in $r=0$, $\theta=\frac{\ \pi}{2}$.} and is ring singularity, while the singularity in $\Delta=0$ exists only if\footnote{This is because the equation $\Delta=0$ admits real solutions only if the discriminant is non-negative, i.e. only if $4m^2-4a^2 \geq 0 \Rightarrow m^2 \geq a^2$.}$M^2 \geq a^2$; in this case it can be shown that, passing to the Kerr coordinates, the singularity disappears and it is therefore a coordinate singularity corresponding to an horizon.\footnote{In reality, since the equation is quadratic there are two event horizons, one internal and external.}.
For the cosmic censorship principle of R. Penrose a gravitational singularity must be hidden by an event horizon and this prompts us to choose $m^2 \geq a^2$ and therefore, with reference to the \ref{Oriz} with $b=0$, you must have $\sigma_2$ real; the consequence is that, from \ref{pop}, the two poles of the dressing matrix must be real. With this choice of poles the \ref{Kerr} is the Kerr metric in the Boyer-Lindquist coordinates\footnote{They are a generalization of the spherical coordinates. The transition from a Cartesian triple ($x^1,x^2,x^3$) to a Boyer-Lindquist one ($r,\theta,\phi$) is given by $x^1=\sqrt{r^2+a^2}sin(\theta)cos(\phi)$,$x^2=\sqrt{r^2+a^2}sin(\theta)sin(\phi)$,$x^3=rcos(\theta)$.} which describes an uncharged rotating black hole with a ratio between mass and angular momentum equal to $a$. If this ratio were null it would mean that the black hole does not rotate, in fact for $a=0$ we obtain the Schwarzschild metric. 

\section{Conclusions}

Gravitational solitons are interesting solutions of the Vacuum field equations. They are rather different to hydrodynamics solitons studied with the KdV example. From the mathematical point of view the gravitational generalization of the spectral transform is not well defined: to study metric in four dimensions we need to fix some components arbitrarily loosing fully generality. This is an open point and we need to develop a method which can be completely general. 

Gravitational solitons have a lot of interesting property, one of the the most interesting is the fusion of poles where different solitons can merge forming solitons associated to poles of higher orders.

Morover, Schwarzschild and Kerr black holes are real two-pole stationary solitons with flat background metrics though at the moment the deep meaning of this conclusion is not completely clear.

The last comment is reserved for the topological structure of the soliton solution leading to black hole metrics. In general, a particular soliton solution that has a different topological structure from that possessed by the void of the theory is called a topological soliton. An example are some solitons that emerge in elementary particle physics, which have a topological structure different from the vacuum which prevents them, even if they are very energetic solutions, from decaying into particles obtained as excitation of the fields from the vacuum. This is because it is assumed that the quantum excitations with respect to the vacuum are associated with a continuous deformation of the fields, a deformation that therefore does not change the topological properties and gives the particles the same topological structure as the vacuum. Although within the theory of gravitation it is not clear when a soliton is a topological object with a topological charge. For Kerr black hole it can be concluded that they are topological solitons: Kerr black holes have a ring-shaped singularity and therefore there exist non-contractable laces; on the other hand in the unperturbed space, i.e. flat space-time, there exist contractable laces. 

\appendix

\section{Multiscale expansion for KdV equation}\label{appA}
Multiscale expansion generalizes the search for a solution through power series expansion by considering slow variables, for example $t_n=\epsilon^nt$ with $\epsilon<<1$. The reason lies in the fact that the simple expansion in power series ceases to be asymptotic, due to the non-linearity effects that generate resonance terms, at a time of the order $t=O(\epsilon^{- 1})$. The advantage of having defined some slow variables is that they allow us to cancel the resonant terms that emerge anyway. The choice and definition of the correct slow variables is recommended to us by the linearized theory together with the physical hypotheses made on the system.
In this appendix we will apply the multiscale expansion to derive the KdV.

Let us consider the class of nonlinear PDEs
\begin{equation}
    u_{,t}+c(u)u_{,z}+A(u)u_{,z,z,z}+B(u)u_{,z}u_{,z,z}+C(u)u_{,z}^3=0,
\end{equation}
with the hypothesis of small amplitudes we can look for a solution as a power series, $u=\sum_n\epsilon^nu_n$with $\epsilon<<1$, obtaining the first order in $\epsilon$ the linear equation
\begin{equation}
\begin{aligned}
u_{1,t}+c(u_0)u_{1,z}+F(u_0)u_{1,z,z,z}=0
\end{aligned}
\end{equation}
which belongs to the class of dispersive equations \footnote{They are those differential equations that admit solutions of the form $Ae^{i\theta}=Ae^{i(\textbf{k}\cdot\textbf{z}-\omega(\textbf{k})t)}$ with $\omega(\textbf{k})\in\textbf{R}$ if$\textbf{k}\in\textbf{R}$ and such that $\nabla^2\omega(\textbf{k})\ne0$. The function $\omega(\textbf{k})$ is called the dispersion relation.} With a dispersion relation equal to $\omega(k)=c(u_0)k-F(u_0)k^3$. Making the further physical hypothesis of long waves we have $\omega(k)\simeq c(u_0)k$ and we find ourselves in the weakly dispersive regime; so let's set $k=\epsilon^{\alpha}q$, then
\begin{equation}
    \theta=kz-\omega(k)t=\epsilon^{\alpha}q(z-c(u_0)t)+F(u_0)\epsilon^{3\alpha}q^3t
\end{equation}
which suggests the definition of the following slow variables $z_1=\epsilon^{\alpha}z$, $t_1=\epsilon^{\alpha}t$ e $t_3=\epsilon^{3\alpha}t$; which implies $\partial_{z}\rightarrow \epsilon^{\alpha} \partial_{z_1}$ and $\partial_{t}\rightarrow \epsilon^{\alpha}\partial_{t_1}+\epsilon^{3\alpha}\partial_{t_3}$.
At order $O(\epsilon^{1+\ alpha})$ we have
\begin{equation}
\begin{aligned}
    &(\epsilon^{\alpha}\partial_{t_1}+\epsilon^{3\alpha}\partial_{t_3})(u_0+\epsilon u_1+..+)+c(u_0)\epsilon^{\alpha}\partial_{z_1}(u_0+\epsilon u_1+..+)+\\
    &+F(u_0)e^{3\alpha}\partial_{z_1}\partial_{z_1}\partial_{z_1}(u_0+\epsilon u_1+...) \Rightarrow u_{,t_1}+c(u_0)u_{1,z_1},
\end{aligned}    
\end{equation}
that is, the advection equation that describes a rigid profile with velocity $c(u_0)$ and the solution is given by $u(z_1-c(u_0)t_1,t_3)$.
For higher orders we have contributions from, in principle, different orders: $O(\epsilon^{2+\alpha})$ and $ O(\epsilon^{1+3\alpha}) $; however, nature favors situations in which the maximum number of terms is balanced at a certain order \footnote {This principle is called the maximum balance.}. Following this principle we have
\begin{equation*}
    2+\alpha=1+3\alpha \Rightarrow \alpha=\frac{1}{2}
\end{equation*}
and therefore the next order is $O(\epsilon^{\frac{5}{2}})$, obtaining 
\begin{equation}
u_{2,t_1}+c(u_0)u_{2,z_1}+\underbrace{u_{1,t_3}+c'(u_0)u_1u_{1,z_1}+F(u_0)u_{1,z_1,z_1,z_1}}_{=h(z_1-c(u_0)t_1,t_3)}=0
\end{equation}
but since the addition of $h(z_1-c(u_0)t_1,t_3)$ is a resonant term we must impose that
\begin{equation}
    u_{1,t_3}+c'(u_0)u_1u_{1,z_1}+F(u_0)u_{1,z_1,z_1,z_1}=0;
\label{KdV}    
\end{equation}
the \ref{KdV} is the equation KdV and fixes the dependence on the variable $t_3$, which we are allowed only because we have defined slow variables and used multiscale expansion; if we had not done this, the resonant term would have destroyed the asymptotic power series. In conclusion, the KdV is the model equation in the description of weakly nonlinear and weakly dispersive systems in the long wave limit.

\section{Check for solutions of equations \ref{zzz}}\label{B}

Let us consider the second of the \ref{zzz} and, replacing the solution \ref{zzzz}, we should have
\begin{equation}
\begin{aligned}
    &\frac{\rho}{\pm\sqrt{(w_k-z)^2+\rho^2}}=\frac{2[(w_k-z)\pm \sqrt{(w_k-z)^2+\rho^2}]\rho}{-2(z-w_k)[(w_k-z)\pm \sqrt{(w_k-z)^2+\rho^2}]+2\rho^2} \Rightarrow\\ &\Rightarrow (w_k-z)[(w_k-z)\pm \sqrt{(w_k-z)^2+\rho^2}]+\rho^2=\\
    &=[(w_k-z)\pm \sqrt{(w_k-z)^2+\rho^2}][\pm\sqrt{(w_k-z)^2+\rho^2}] \Rightarrow \\
    & \Rightarrow (w_k-z)^2\pm (w_k-z)\sqrt{(w_k-z)^2+\rho^2}+\rho^2=\\
    &=\pm (w_k-z)\sqrt{(w_k-z)^2+\rho^2}+(w_k-z)^2+\rho^2 \Rightarrow 0=0
\end{aligned}    
\end{equation}
and the check is complete. For the second of the \ref{zzz} the procedure is the same.

\end{document}